\documentclass{article}

\usepackage{pcf_nat}

\usepackage[utf8]{inputenc} % allow utf-8 input
\usepackage[T1]{fontenc}    % use 8-bit T1 fonts
\usepackage{hyperref}       % hyperlinks
\usepackage{url}            % simple URL typesetting
\usepackage{booktabs}       % professional-quality tables
\usepackage{amsfonts}       % blackboard math symbols
\usepackage{nicefrac}       % compact symbols for 1/2, etc.
\usepackage{microtype}      % microtypography
\usepackage{xcolor}         % colors
\usepackage{graphicx}
\usepackage{amsmath}

\title{Neighborhood Attention Transformer with Progressive Channel Fusion for Speaker Verification}

\author{%
  Nian Li, Jianguo Wei\thanks{Corresponding author.} \\
  College of Intelligence and Computing, Tianjin University, Tianjin, China \\
  \texttt{\{linian1996, jianguo\}@tju.edu.cn} \\
}

\begin{document}

\maketitle

\begin{abstract}
Transformer-based architectures for speaker verification typically require more training data than ECAPA-TDNN. Therefore, recent work has generally been trained on VoxCeleb1\&2. We propose a backbone network based on self-attention, which can achieve competitive results when trained on VoxCeleb2 alone. The network alternates between neighborhood attention and global attention to capture local and global features, then aggregates features of different hierarchical levels, and finally performs attentive statistics pooling. Additionally, we employ a progressive channel fusion strategy to expand the receptive field in the channel dimension as the network deepens. We trained the proposed PCF-NAT model on VoxCeleb2 and evaluated it on VoxCeleb1 and the validation sets of VoxSRC. The EER and minDCF of the shallow PCF-NAT are on average more than 20\% lower than those of similarly sized ECAPA-TDNN. Deep PCF-NAT achieves an EER lower than 0.5\% on VoxCeleb1-O. The code and models are publicly available at \url{https://github.com/ChenNan1996/PCF-NAT}.
\end{abstract}

\section{Introduction}

Automatic Speaker Verification (ASV) is a task that determines whether two utterances belong to the same speaker. As an important method of biometric identification, it finds wide applications in areas such as door locks, phone unlocking, and conference recording. In recent years, mainstream ASV systems have all adopted deep neural networks. During training, a classification task is performed, consisting of an embedding extractor and a classifier. During application, the embedding extractor is used to extract embeddings from two utterances, and then the similarity between these embeddings is calculated as the probability that the two utterances belong to the same speaker. The speaker embeddings extracted by ASV can be used for downstream tasks such as speech synthesis and voice conversion.

Based on network structure, neural network-based ASV systems mainly fall into three types: Time-Delay Neural Networks (TDNN, implemented using one-dimensional convolution), two-dimensional convolutional neural networks, and Transformer-based neural networks.

X-vector~\cite{x-vector} was the first to employ TDNN for speaker verification, and it projected variable-length inputs into fixed-length representations through statistical pooling. Recently, ECAPA-TDNN~\cite{ECAPA-TDNN} has made impressive advancements over X-vector. ECAPA-TDNN aggregates multi-level features and utilizes attentive statistics pooling. Subsequent works have further improved upon ECAPA-TDNN. PCF-ECAPA-TDNN~\cite{PCF-ECAPA-TDNN} adopts a progressive channel fusion strategy, enabling the receptive field in the channel dimension to expand as the network deepens, leading to better performance. However, ECAPA-TDNN and its variants struggle to improve performance by deepening the network or increasing channels.

Therefore, the network structure based on two-dimensional convolution~\cite{zeinali2019description}~\cite{thienpondt2020idlab}~\cite{zhao2021speakin}~\cite{makarov2022id}~\cite{torgashov2023id}~\cite{zheng2023unisound} is widely used in The VoxCeleb Speaker Recognition Challenge (VoxSRC)~\cite{VoxSRC2019}~\cite{VoxSRC2020}~\cite{VoxSRC2021}~\cite{VoxSRC2022}. Although the effectiveness of two-dimensional convolutional network structures is not comparable to ECAPA-TDNN and its variants in small-scale scenarios, they offer greater scalability. The two-dimensional convolution-based network structures currently used in VoxSRC are variants of Resnet100+ or even Resnet200+~\cite{He_2016_CVPR}. Additionally, research indicates that using two-dimensional convolution followed by ECAPA-TDNN yields better results~\cite{thienpondt21_interspeech}~\cite{wang23ha_interspeech}.

Recently, Transformer-based~\cite{transformer} architectures have been explored for speaker embedding extraction. Transformers excel in modeling long-range global context and facilitate efficient parallel training. However, many existing studies indicate that without complex pre-training procedures and large parameters, Transformers struggle to achieve satisfactory performance in ASV~\cite{wang2021unispeech}~\cite{9814838}~\cite{9747814}.

The MFA-Conformer~\cite{MFA-Conformer} utilizes the conformer architecture~\cite{gulati20_interspeech}, which combines convolution and self-attention, and adopts a network structure similar to ECAPA-TDNN, thereby enhancing the effectiveness of Transformer. Nevertheless, some studies~\cite{10096659} suggest that MFA-Conformer is prone to overfitting due to limited speaker recognition training data, and its performance on VoxCeleb2~\cite{chung18b_interspeech} is inferior to ECAPA-TDNN. Additionally, attempts to introduce SwinTransformer~\cite{SwinTransformer} from computer vision into ASV have yielded unsatisfactory results~\cite{10096333}.

This paper introduces Neighborhood Attention~\cite{NAT}, a concept from computer vision, into ASV. We alternate between neighborhood attention and global attention (original self-attention) to capture local and global features. Adopting an architecture similar to ECAPA-TDNN, we concatenate the outputs of all blocks along the channel dimension and then apply attentive statistics pooling. Additionally, We use the progressive channel fusion strategy proposed by PCF-ECAPA-TDNN, replacing linear layers with 1-dimensional group convolutions with a kernel size of 1, and gradually reducing the number of groups to expand the receptive field in the channel dimension.

The organization of this paper is as follows: Section 2 describes related work and serves as the baseline. Section 3 details the proposed network architecture and its components. Section 4 outlines the experimental setup, Section 5 discusses the experimental results, and Section 6 concludes the paper and outlines future work.

\section{Related Work}
\label{gen_inst}

\subsection{ECAPA-TDNN}

ECAPA-TDNN employs three 1-Dimensional Squeeze-Excitation Res2Blocks and gradually increases the dilation of dilated convolutions to expand the receptive field. The Squeeze-Excitation mechanism scales frame-level features based on the global properties of the input along the time dimension. The Res2Block divides the input channels into several segments and performs convolution sequentially (the result of the previous convolution is added to the input as the input for the subsequent convolution) and then concatenates the results to capture multi-scale features.

Believing that shallow features can also contribute towards more robust speaker embeddings, ECAPA-TDNN concatenates the outputs of all SE-Res2Blocks along the channel dimension, and then passes them through a dense layer before feeding them into the pooling layer. Another contribution of ECAPA-TDNN is attentive statistics pooling (ASP), which extends the attention mechanism from the time dimension to the channel dimension. This enables the network to focus more on speaker characteristics that are not activated at similar times.

\subsection{PCF-ECAPA-TDNN}

PCF-ECAPA-TDNN uses group 1D convolutions and gradually reduces the number of groups, thereby expanding the receptive field in the frequency (channel) dimension as the network deepens. This is known as the progressive channel fusion strategy. A key difference between ResNet and TDNN is that 2D convolutions in ResNet have a local receptive field in both the frequency and time dimensions, whereas TDNN maintains a global receptive field in the frequency dimension throughout, which increases the risk of overfitting due to the large number of parameters. Additionally, PCF-ECAPA-TDNN deepens the network by using four blocks, each containing two SE-Res2Blocks. Unlike the Res2Block in ECAPA-TDNN, PCF-ECAPA-TDNN adds a branch with a kernel size of 1.

\subsection{MFA-Conformer}

MFA-Conformer applies the Conformer, which uses convolutions to capture local features and self-attention to capture global features, to speaker verification. Unlike the previous two baseline models, MFA-Conformer performs down-sampling in the time dimension. After the down-sampling layer, there are multiple Conformer blocks. The optimal configuration is a down-sampling rate of 1/2 and six blocks. Specifically, this model downsamples Fbank features using a 2D convolution with a kernel size of 3 and a stride of 2, and adjusts the number of channels to 256 through a linear layer. Each Conformer block contains two Macaron-like feed-forward modules with half residual connections, sandwiching the multi-head self-attention and convolution modules. Similar to ECAPA-TDNN, MFA-Conformer also employs multi-layer feature aggregation and uses attentive statistics pooling.

\section{Proposed Architecture}
\label{headings}

\begin{figure}
  \centering
   \includegraphics[height=13.5cm, width=12.39cm]{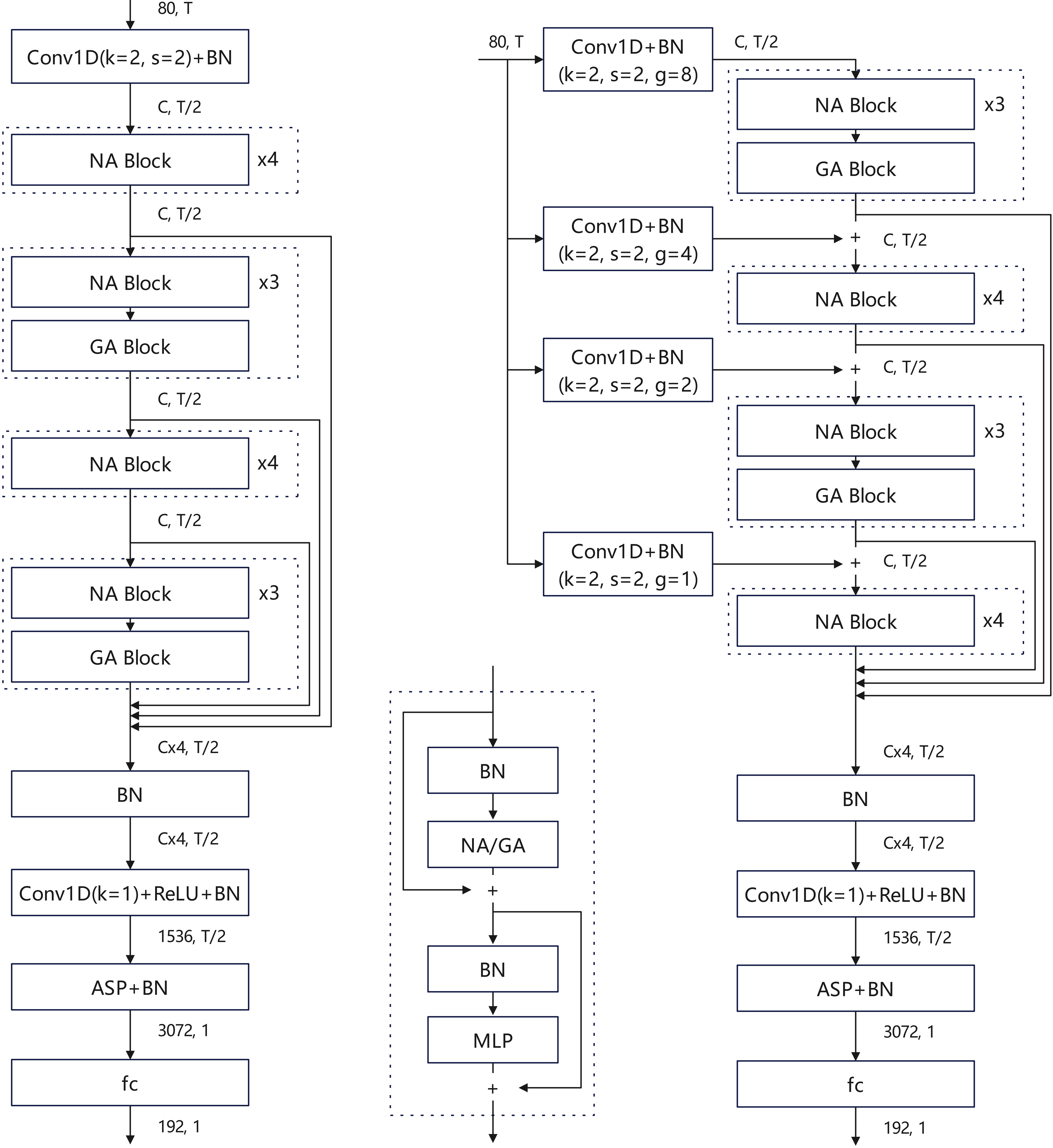}
  \caption{Proposed MFA-NAT (left) and PCF-NAT (right). The structure at the bottom-middle of the figure is the NA/GA block.}
\end{figure}

We propose an architecture that alternates between neighborhood attention (NA) and global attention (GA) to capture both local and global features. Specifically, we only employ global attention in the final layer of the 2nd and 4th blocks. Similar to ECAPA-TDNN, we also utilize multi-level feature aggregation (MFA) and attentive statistics pooling (ASP). The architecture overview of MFA-NAT is provided in the left part of Figure 1. Furthermore, we introduce a progressive channel fusion strategy to enhance performance. There are some structural changes in PCF-NAT compared to MFA-NAT, as illustrated in the right part of Figure 1.

Additionally, the downsampling layer includes a one-dimensional convolution with a kernel size of 2 and a stride of 2, along with a normalization layer. To maintain consistency with the pooling layer, we utilize batch normalization as the normalization layer throughout the entire model.

\subsection{Neighborhood Attention}

\begin{figure}
  \centering
   \includegraphics[height=4.21cm, width=13cm]{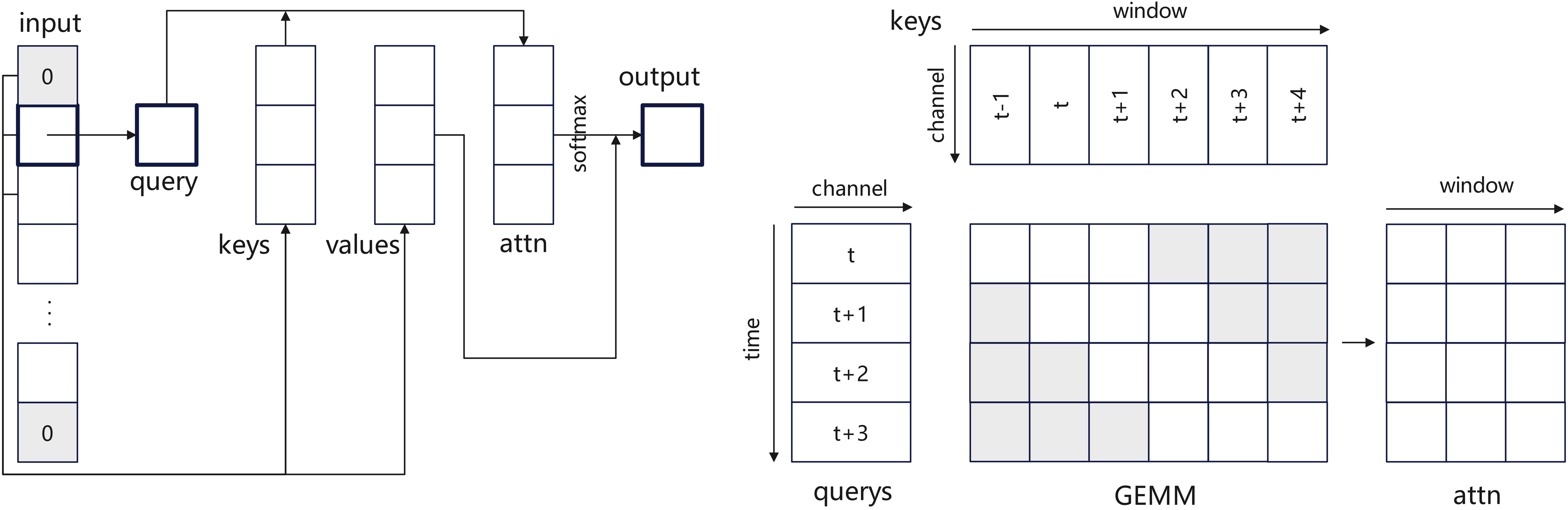}
  \caption{An example of a NA with window size of 3. The right part of the figure is the process of computing the attention matrix using the GEMM approach.}
\end{figure}

Self-attention in the original Transformer exhibits quadratic complexity concerning the number of tokens, posing challenges for processing high-resolution images or long speech. Moreover, self-attention lacks certain inductive biases, such as the locality inherent in convolution, which are essential for effective learning without massive amounts of data or advanced training techniques and augmentations~\cite{dosovitskiy2020image}~\cite{hassani2021escaping}~\cite{touvron2021training}. 

To address these issues, window attention has emerged as a promising solution. In recent years, the SwinTransformer~\cite{SwinTransformer} and Neighborhood Attention Transformer (NAT)~\cite{NAT} have sequentially demonstrated outstanding performance in visual tasks. SwinTransformer partitions an image into multiple patches, applying self-attention to non-overlapping windows and shifting windows to expand the receptive field. On the other hand, Stand-Alone Self-Attention (SASA)~\cite{ramachandran2019stand} ensures each pixel attends to a window around it, akin to convolutions. Compared to shift window attention, SASA offers a more consistent receptive field and quicker expansion with network depth. However, SASA's speed was limited due to the lack of an efficient implementation similar to that of convolutions.

Neighborhood Attention (NA) shares similarities with SASA but restricts the window from extending beyond the image boundary, eliminating surrounding padding. Additionally, the authors of NA developed an efficient implementation of NA~\cite{hassani2024faster} and used a network structure in Neighborhood Attention Transformer (NAT) similar to that of the SwinTransformer.

As illustrated in the left part of Figure 2, our model follows the approach of SASA, padding the ends of the sequence, although we still refer to it as NA. Given a window size \( W \), and the Query matrix \( Q \), Key matrix \( K \), and Value matrix \( V \), the method for calculating the attention matrix is:
\begin{equation}
   attn_{t,w}=\sum\nolimits_{c=0}^{C}{Q_{t,c}K_{t-\lfloor W/2 \rfloor +w,c}}.
   \label{attention matrix}
\end{equation}
The attention matrix then undergoes a softmax operation. Finally, the formula for computing the output is as follows:
\begin{equation}
   output_{t,c}=\sum\nolimits_{w=0}^{W}{attn_{t,w}V_{t-\lfloor W/2 \rfloor +w,c}}.
   \label{output}
\end{equation}

We have also developed a Python package for NA with efficient C++ and CUDA kernels. To leverage the Tensor Cores of Nvidia GPUs, we convert vector-matrix multiplication into matrix-matrix multiplication, even though this introduces some redundant computations, as illustrated on the right side of Figure 2. Similar to computing the attention matrix, the gradients for the Query matrix and the Key matrix are also computed using the GEMM (General Matrix Multiply) approach. As for computing the output, calculating the gradients of the attention matrix and the Value matrix, the processes can be reused from the previous steps. For an \( MNK \) matrix multiplication, with a window size \( W \), the proportion of effective computation can be determined as follows:
\begin{equation}
   ratio=\frac{W}{\lceil(M+W-1)/N\rceil \cdot N}.
   \label{effective computation ratio}
\end{equation}
In our implementation, we use a \( M16N8K16 \) matrix multiplication with a window size of 27, resulting in an effective computation ratio of 56.26\%. Although this ratio is not particularly high, the significant speed advantage of Tensor Cores over CUDA Cores means that converting GEMV to GEMM still yields more than a 5x acceleration.

\subsection{Progressive Channel Fusion}

To ensure that the receptive field in the channel dimension expands progressively with network depth, similar to the time dimension, and to reduce the number of parameters to mitigate the risk of overfitting, we use a large number of 1D group convolutions, similar to PCF-ECAPA-TDNN~\cite{PCF-ECAPA-TDNN}. Specifically, we use four 1D group convolutions to process the input Fbank features, with the number of groups set to 8, 4, 2, and 1, respectively, and then feed the results into four blocks. The input to the second and subsequent blocks is augmented with the output from the previous block. For the four blocks, we replace the linear layers, including the \(qkv\) layers, with 1D group convolutions, with the number of groups also set to 8, 4, 2, and 1. Unlike PCF-ECAPA-TDNN, we reduce the computational load by down-sampling the Fbank features in the time dimension by half, setting both the kernel size and stride of the convolution to 2.

\section{Experiments}
\label{others}

\subsection{Dataset}

We utilized voxceleb2-dev~\cite{chung18b_interspeech} as the training set, comprising 1,092,009 utterances from 5,994 speakers. To augment the data, each utterance was sped up or down by a factor of 0.9 or 1.1, resulting in shifted pitch utterances treated as from new speakers~\cite{yamamoto2019speaker}~\cite{ko2015audio}~\cite{wang2020dkudukeece}. Ultimately, the training data contained 3,276,027 utterances from 17,982 speakers. Additionally, we employed on-the-fly data augmentation to introduce additive background noise or convolutional reverberation noise for the time-domain waveform using MUSAN (music, speech and noise)~\cite{snyder201musan} and RIRS-NOISES~\cite{ko2017study}. For evaluation, we use the official evaluation sets including VoxCeleb1-O, VoxCeleb1-E, and VoxCeleb1-H~\cite{nagrani17_interspeech}. To more accurately assess the generalization ability of the models, we also use the validation sets from the last four years of the VoxSRC~\cite{VoxSRC2020}~\cite{VoxSRC2021}~\cite{VoxSRC2022}.

\subsection{Model Settings}

For fair comparisons, we standardized the output channels of the MFA module to 1536 to match the original configuration of ECAPA-TDNN. The speaker embedding dimension for all systems was set to 192.

\paragraph{ECAPA-TDNN.}

We used the ECAPA-TDNN implemented by SpeechBrain~\cite{ravanelli2021speechbrain} as the first baseline model, which includes three SE-Res2Blocks with 512 channels. These three blocks use 1D dilated convolutions with dilations of 2, 3, and 4, respectively.

\paragraph{PCF-ECAPA-TDNN.}

Following the original configuration, this model includes four blocks, each containing two SE-Res2Blocks with branches and 512 channels. The numbers of groups for the 1D group convolutions in the blocks are 8, 4, 2, and 1, respectively.

\paragraph{MFA-Conformer.}

We utilized the source code provided by the authors and adopted the best-performing model configuration, including a 1/2 subsampling rate and 6 Conformer blocks with 256 channels. For the multi-head self-attention modules in the Conformer blocks, the number of attention heads is set to 4. For the 1D convolution modules, the kernel size is set to 15. For the feed-forward modules, the number of units in the linear hidden layer is set to 2048.

\paragraph{MFA-NAT.}

Similar to PCF-ECAPA-TDNN, our proposed model consists of four blocks. The first and third blocks use neighborhood attention entirely, while the second and fourth blocks only use global attention in the last layer. Similar to MFA-Conformer, we use only four heads in the global attention layers to reduce memory usage and set the channel number to 256 for all blocks. For the neighborhood attention layers, the number of heads is set to a more efficient 16, with a window size of 27. Additionally, we scale the model by adjusting the number of attention layers in each block. For example, MFA-NAT (3$\times$4) indicates that each block in the model contains three attention layers. Finally, the drop path rate for MFA-NAT models of different depths is set to 1.0.

\paragraph{PCF-NAT.}

The proposed model includes four pairs of down-sampling layers and blocks, with the number of groups set to 8, 4, 2, and 1, respectively, to be consistent with PCF-ECAPA-TDNN. Unlike MFA-NAT, the two global attention layers of PCF-NAT are placed at the end of the first and third blocks. Additionally, PCF-NAT models of different depths use different drop path rates: PCF-NAT (3$\times$4), (4$\times$4), (5$\times$4), and (6$\times$4) use drop path rates of 1.0, 1.1, 1.2, and 1.3, respectively. For other configurations, PCF-NAT is identical to MFA-NAT.

\subsection{Training}

Fixed-length 3-second segments were randomly extracted from each utterance, with longer utterances sampled multiple times in one epoch. The input features consisted of 80-dimensional log-Mel Filter Banks with a window length of 25 ms and a frame-shift of 10 ms. FBank features were normalized through cepstral mean subtraction, and no voice activity detection was applied. 

A mini-batch contained 256 original utterances and 256 augmented utterances. All models were trained using AAM+K-subcenter Softmax loss~\cite{deng2019arcface}~\cite{deng2020sub} with a margin of 0.2, a scaling factor of 32, and k of 3. The SGD optimizer~\cite{sutskever2013importance} was employed with an initial learning rate of 0.0001, increased to 0.5 at the end of epoch 1. Subsequently, \(CosineAnnealingLR\)~\cite{loshchilov2016sgdr} was used to reduce the learning rate to 0.0001 over the next 9 epochs. 

To prevent overfitting, weight decay was applied to all models. Specifically, the weight decay of ECAPA-TDNN, PCF-ECAPA-TDNN, and MFA-Conformer was set to 2e-5, 5e-5, and 1.5e-6, respectively, to maintain consistency with the original configuration. For MFA-NAT and PCF-NAT, the weight decay was set to 1e-5.

\subsection{Evaluation}

To address differences in segment duration between training and evaluation, testing utterances exceeding 8 seconds were divided into segments of 4 to 6 seconds. The embeddings of these segments were normalized, and the mean was calculated as the embedding for the utterances~\cite{cai2022kriston}.

Trial scores were produced using the inner product between embeddings, followed by normalization using adaptive S-norm~\cite{karam2011towards}~\cite{matejka2017analysis}. The imposter cohort comprised speaker-wise averages of length-normalized embeddings of all training utterances, with the size set to 300 for VoxCeleb1-O, VoxCeleb1-E, VoxSRC2022, and VoxSRC2023, and to 100 for VoxCeleb1-H, VoxSRC2020, and VoxSRC2021.

Equal Error Rate (EER) and minimum Detection Cost Function (minDCF) were reported with CFA = CMiss = 1 for performance evaluation. The Ptarget of minDCF was set to 0.01 for VoxCeleb1 and 0.05 for VoxSRC. Furthermore, we calculate the number of batches processed per second by the models on a single Nvidia GPU 4090 to evaluate the inference speeds. Each batch contains 512 six-second utterances. We use PyTorch's \(torch.compile\) with the default mode to accelerate model inference.

\section{Results and Discussions}

\subsection{Results on VoxCeleb1 and VoxSRC}

\begin{table}
  \caption{EER and minDCF$_{0.01}$ performance of all systems on VoxCeleb1.}
  \label{results-table1}
  \centering
  \begin{tabular}{lrrrcccccc}
    \toprule
    & & & \multicolumn{2}{c}{Vox1-O} & \multicolumn{2}{c}{Vox1-E} & \multicolumn{2}{c}{Vox1-H}  \\
    \cmidrule(r){4-5} \cmidrule(r){6-7} \cmidrule(r){8-9}
    System             & Params & Bs/s & EER & DCF & EER & DCF & EER & DCF \\
    \midrule
    MFA-Conformer      & 20.5M & 7.03  & 0.691  & 0.1091  & 0.893  & 0.0997  & 1.709  & 0.1654  \\
    ECAPA-TDNN         & 14.7M  & 9.89  & 0.681  & 0.1068  & 0.871  & 0.0961  & 1.643  & 0.1672  \\
    PCF-ECAPA          & 8.9M   & 7.93  & 0.590  & 0.0667  & 0.758  & 0.0877  & 1.453  & 0.1455  \\
    \midrule
    MFA-NAT (3$\times$4)       & 12.6M  & 10.34  & 0.558  & 0.0731  & 0.738  & 0.0818  & 1.347  & 0.1328  \\
    MFA-NAT (4$\times$4)       & 15.8M  & 8.32  & 0.580  & 0.0601  & 0.711  & 0.0777  & 1.305  & 0.1295  \\
    MFA-NAT (5$\times$4)       & 18.9M  & 6.96  & 0.542  & 0.0792  & 0.721  & 0.0752  & 1.327  & 0.1351  \\
    MFA-NAT (6$\times$4)       & 22.1M  & 5.98  & 0.601  & 0.0606  & 0.718  & 0.0749  & 1.305  & 0.1340  \\
    \midrule
    PCF-NAT (3$\times$4)       & 7.6M  & 8.52  & 0.537  & 0.0500  & 0.703  & 0.0790  & 1.304  & 0.1332  \\
    PCF-NAT (4$\times$4)       & 9.0M  & 6.82  & 0.526  & 0.0604  & 0.672  & 0.0750  & 1.260  & 0.1298  \\
    PCF-NAT (5$\times$4)       & 10.5M  & 5.69  & 0.452  & \textbf{0.0392}  & 0.672  & 0.0715  & 1.236  & \textbf{0.1270}  \\
    PCF-NAT (6$\times$4)       & 12.0M  & 4.87  & \textbf{0.436}  & 0.0467  & \textbf{0.654}  & \textbf{0.0712}  & \textbf{1.198}  & 0.1284  \\
    \bottomrule
  \end{tabular}
\end{table}

\begin{table}
  \caption{EER and minDCF$_{0.05}$ performance of all systems on the VoxSRC validation sets.}
  \label{results-table2}
  \centering
  \begin{tabular}{lcccccccc}
    \toprule
    & \multicolumn{2}{c}{VoxSRC2020} & \multicolumn{2}{c}{VoxSRC2021} & \multicolumn{2}{c}{VoxSRC2022} & \multicolumn{2}{c}{VoxSRC2023}  \\
    \cmidrule(r){2-3} \cmidrule(r){4-5} \cmidrule(r){6-7} \cmidrule(r){8-9}
    System             & EER & DCF & EER & DCF & EER & DCF & EER & DCF \\
    \midrule
    MFA-Conformer      & 2.812  & 0.1499  & 3.290  & 0.1789  & 2.277  & 0.1517  & 4.075  & 0.2323 \\
    ECAPA-TDNN         & 2.747  & 0.1460  & 3.537  & 0.1861  & 2.224  & 0.1559  & 4.133  & 0.2373 \\
    PCF-ECAPA          & 2.493  & 0.1311  & 3.067  & 0.1677  & 1.981  & 0.1362  & 3.775  & 0.2130 \\
    \midrule
    MFA-NAT (3$\times$4)       & 2.272  & 0.1214  & 2.671  & 0.1446  & 1.862  & 0.1260  & 3.447  & 0.1923 \\
    MFA-NAT (4$\times$4)       & 2.229  & 0.1177  & 2.606  & 0.1453  & 1.815  & 0.1187  & 3.357  & 0.1845 \\
    MFA-NAT (5$\times$4)       & 2.255  & 0.1190  & 2.634  & 0.1547  & 1.833  & 0.1241  & 3.334  & 0.1949 \\
    MFA-NAT (6$\times$4)       & 2.232  & 0.1185  & 2.716  & 0.1588  & 1.764  & 0.1202  & 3.328  & 0.1852 \\
    \midrule
    PCF-NAT (3$\times$4)       & 2.202  & 0.1209  & 2.609  & 0.1423  & 1.793  & 0.1262  & 3.340  & 0.1938 \\
    PCF-NAT (4$\times$4)       & 2.163  & 0.1153  & 2.456  & \textbf{0.1403}  & \textbf{1.723}  & 0.1170  & 3.244  & \textbf{0.1830} \\
    PCF-NAT (5$\times$4)       & 2.139  & 0.1120  & 2.723  & 0.1509  & 1.724  & 0.1164  & 3.307  & 0.1843 \\
    PCF-NAT (6$\times$4)       & \textbf{2.059}  & \textbf{0.1074}  & \textbf{2.426}  & 0.1405  & 1.723  & \textbf{0.1153}  & \textbf{3.186}  & 0.1830 \\
    \bottomrule
  \end{tabular}
\end{table}

In this section, we present the performance of the proposed MFA-NAT and PCF-NAT with different depths, as well as the performance of the three baseline systems: MFA-Conformer, ECAPA-TDNN, and PCF-ECAPA-TDNN. Table 1 reports the EER and minDCF on VoxCeleb1, along with the number of model parameters in the embedding extractor and the number of batches processed per second during inference. The EER and minDCF on the validation sets of VoxSRC are provided in Table 2.

As shown in the two tables, ECAPA-TDNN's EER and minDCF are close to those of MFA-Conformer, but ECAPA-TDNN has fewer parameters and faster inference speed. Among the three baseline models, PCF-ECAPA-TDNN achieves the lowest EER and minDCF with the fewest parameters, but its throughput during inference is lower than the shallower ECAPA-TDNN.

For scenarios requiring fast inference, the proposed MFA-NAT (3$\times$4) is a good choice. Its throughput is 4.55\% higher than ECAPA-TDNN, and its EER and minDCF are 6.86\% lower on average than those of PCF-ECAPA-TDNN. By deepening the network, MFA-NAT (4$\times$4) achieves EER and minDCF that are 3.57\% lower on average than MFA-NAT (3$\times$4), although its throughput is reduced by 19.54\%. Further deepening of MFA-NAT does not yield better performance.

The use of the progressive channel fusion strategy allows PCF-NAT to achieve better performance and scalability. Compared to the EER and minDCF of MFA-NAT (4$\times$4), PCF-NAT (4$\times$4) shows an average reduction of 3.28\%, while the deeper PCF-NAT (6$\times$4) achieves an average reduction of 8.07\%. However, in the current Pytorch implementation, 1D group convolutions are slower than linear layers despite having fewer parameters and computations. This results in less satisfactory inference speed for PCF-NAT, which uses many 1D group convolutions.

Not mentioned in the tables but equally important is that PCF-NAT uses over 25\%, 40\%, and 45\% less peak memory during inference compared to MFA-Conformer, ECAPA-TDNN, and PCF-ECAPA-TDNN, respectively. MFA-NAT and PCF-NAT have similar peak memory usage during inference. Additionally, because many intermediate values do not need to be saved during inference, deepening these two models does not increase memory usage. This makes MFA-NAT or PCF-NAT suitable for environments with limited memory capacity.

\subsection{Ablation Study}

\begin{table}
  \caption{Ablation study of MFA-NAT (4$\times$4). DCF is minDCF$_{0.01}$. w/o is without. NA is neighborhood attention. GA is global attention.}
  \label{ablation-study-table}
  \centering
  \begin{tabular}{lrccccccr}
    \toprule
    & &  \multicolumn{2}{c}{Vox1-O} & \multicolumn{2}{c}{Vox1-E} & \multicolumn{2}{c}{Vox1-H} &  \\
    \cmidrule(r){3-4} \cmidrule(r){5-6} \cmidrule(r){7-8}
    No. & System             & EER & DCF & EER & DCF & EER & DCF & avg$\uparrow$  \\
    \midrule
    0. & MFA-NAT (4$\times$4)           & 0.580  & 0.0601  & 0.711  & 0.0777  & 1.305  & 0.1295  & \\
    \midrule
    1. & w/o Fbank norm      & 0.638  & 0.0858  & 0.751  & 0.0788  & 1.372  & 0.1400  & 12.17\%  \\
    2. & w/o NA padding      & 0.569  & 0.0573  & 0.720  & 0.0785  & 1.323  & 0.1374  & 0.54\%  \\
    3. & w/o GA     & 0.681  & 0.0687  & 0.775  & 0.0830  & 1.402  & 0.1424  & 10.82\%  \\
    4. & with four GAs   & 0.564  & 0.0755  & 0.751  & 0.0821  & 1.343  & 0.1354  & 6.94\%  \\
    5. & with LayerNorm           & 0.537  & 0.0731  & 0.705  & 0.0760  & 1.309  & 0.1353  & 2.66\%  \\
    6. & w/o drop path       & 0.548  & 0.0673  & 0.730  & 0.0833  & 1.351  & 0.1346  & 3.97\%  \\
    7. & w/o MFA             & 0.612  & 0.0712  & 0.713  & 0.0790  & 1.296  & 0.1336  & 4.74\%  \\
    8. & w/o ASP             & 0.633  & 0.0754  & 0.797  & 0.0883  & 1.472  & 0.1465  & 14.38\%  \\
    9. & w/o AS-norm         & 0.669  & 0.0719  & 0.747  & 0.0824  & 1.393  & 0.1360  & 9.64\%  \\
    \bottomrule
  \end{tabular}
\end{table}

The effectiveness of the channel fusion strategy has been explained in the previous subsection. To understand the effects of other components used in the proposed model, We conducted an ablation study on MFA-NAT (4$\times$4) and evaluated the results on VoxCeleb1, as shown in Table 3. In Experiment 2, we did not pad the sequence ends and restricted the neighborhood attention window to within the sequence range, which slightly increased the EER and minDCF. This supports our decision to follow the earlier Stand-Alone Self-Attention approach. In Experiment 3, where only neighborhood attention was used, the EER and minDCF increased by an average of 10.82\%, indicating the usefulness of the two global attention layers in MFA-NAT. However, Experiment 4 showed that adding more global attention layers worsened the results, even if only the last layer of blocks 1 and 3 was changed to global attention. Experiment 5 replaced all batch normalization layers before multi-level feature aggregation with layer normalization, resulting in an average increase of 2.66\% in EER and minDCF. In Experiment 7, removing MFA but retaining its subsequent dense layer led to a deterioration of 4.74\% in the results.

\section{Conclusion and Future Work}

In this paper, we present PCF-NAT, a novel Transformer-based speaker embedding extractor for speaker verification. Unlike Conformer, we capture local features using neighborhood attention instead of convolution, and we employ less global attention. Ablation experiments demonstrate the effectiveness of this alternating use of neighborhood and global attention. PCF-NAT, trained solely on VoxCeleb2, significantly outperforms ECAPA-TDNN without requiring a larger dataset. Additionally, PCF-NAT is highly scalable: its accuracy can be improved by deepening the network, and its inference speed can be increased by omitting the progressive channel fusion strategy. We anticipate that this scalability will facilitate the use of PCF-NAT in downstream tasks such as speech synthesis and voice conversion. For future work, we aim to find more effective down-sampling methods as we are concerned that the current simplistic down-sampling approach may limit the performance of the proposed model. Additionally, we plan to train PCF-NAT on larger datasets to evaluate whether its generalization ability continues to improve with increasing training data.

\newpage
\bibliographystyle{unsrt}
\bibliography{pcf_nat}

\begin{thebibliography}{10}

\bibitem{x-vector}
David Snyder, Daniel Garcia-Romero, Gregory Sell, Daniel Povey, and Sanjeev Khudanpur.
\newblock X-vectors: Robust {DNN} embeddings for speaker recognition.
\newblock In {\em IEEE ICASSP}, pages 5329--5333, 2018.

\bibitem{ECAPA-TDNN}
Brecht Desplanques, Jenthe Thienpondt, and Kris Demuynck.
\newblock {ECAPA-TDNN}: Emphasized channel attention, propagation and aggregation in {TDNN} based speaker verification.
\newblock In {\em Proc. Interspeech}, pages 3830--3834, 2020.

\bibitem{PCF-ECAPA-TDNN}
Zhenduo Zhao, Zhuo Li, Wenchao Wang, and Pengyuan Zhang.
\newblock Pcf: Ecapa-tdnn with progressive channel fusion for speaker verification.
\newblock In {\em ICASSP 2023 - 2023 IEEE International Conference on Acoustics, Speech and Signal Processing (ICASSP)}, pages 1--5, 2023.

\bibitem{zeinali2019description}
Hossein Zeinali, Shuai Wang, Anna Silnova, Pavel Matějka, and Oldřich Plchot.
\newblock But system description to voxceleb speaker recognition challenge 2019, 2019.

\bibitem{thienpondt2020idlab}
Jenthe Thienpondt, Brecht Desplanques, and Kris Demuynck.
\newblock The idlab voxceleb speaker recognition challenge 2020 system description, 2020.

\bibitem{zhao2021speakin}
Miao Zhao, Yufeng Ma, Min Liu, and Minqiang Xu.
\newblock The speakin system for voxceleb speaker recognition challange 2021, 2021.

\bibitem{makarov2022id}
Rostislav Makarov, Nikita Torgashov, Alexander Alenin, Ivan Yakovlev, and Anton Okhotnikov.
\newblock Id r\&d system description to voxceleb speaker recognition challenge 2022.
\newblock {\em ID R\&D Inc.: New York, NY, USA}, 2022.

\bibitem{torgashov2023id}
Nikita Torgashov, Rostislav Makarov, Ivan Yakovlev, Pavel Malov, Andrei Balykin, and Anton Okhotnikov.
\newblock The id r\&d voxceleb speaker recognition challenge 2023 system description, 2023.

\bibitem{zheng2023unisound}
Yu~Zheng, Yajun Zhang, Chuanying Niu, Yibin Zhan, Yanhua Long, and Dongxing Xu.
\newblock Unisound system for voxceleb speaker recognition challenge 2023, 2023.

\bibitem{VoxSRC2019}
Joon~Son Chung, Arsha Nagrani, Ernesto Coto, Weidi Xie, Mitchell McLaren, Douglas~A Reynolds, and Andrew Zisserman.
\newblock Voxsrc 2019: The first voxceleb speaker recognition challenge, 2019.

\bibitem{VoxSRC2020}
Arsha Nagrani, Joon~Son Chung, Jaesung Huh, Andrew Brown, Ernesto Coto, Weidi Xie, Mitchell McLaren, Douglas~A Reynolds, and Andrew Zisserman.
\newblock Voxsrc 2020: The second voxceleb speaker recognition challenge, 2020.

\bibitem{VoxSRC2021}
Andrew Brown, Jaesung Huh, Joon~Son Chung, Arsha Nagrani, and Andrew Zisserman.
\newblock Voxsrc 2021: The third voxceleb speaker recognition challenge.
\newblock {\em arXiv preprint arXiv:2201.04583}, 2022.

\bibitem{VoxSRC2022}
Jaesung Huh, Andrew Brown, Jee weon Jung, Joon~Son Chung, Arsha Nagrani, Daniel Garcia-Romero, and Andrew Zisserman.
\newblock Voxsrc 2022: The fourth voxceleb speaker recognition challenge, 2023.

\bibitem{He_2016_CVPR}
Kaiming He, Xiangyu Zhang, Shaoqing Ren, and Jian Sun.
\newblock Deep residual learning for image recognition.
\newblock In {\em Proceedings of the IEEE Conference on Computer Vision and Pattern Recognition (CVPR)}, June 2016.

\bibitem{thienpondt21_interspeech}
Jenthe Thienpondt, Brecht Desplanques, and Kris Demuynck.
\newblock {Integrating Frequency Translational Invariance in TDNNs and Frequency Positional Information in 2D ResNets to Enhance Speaker Verification}.
\newblock In {\em Proc. Interspeech 2021}, pages 2302--2306, 2021.

\bibitem{wang23ha_interspeech}
Hui Wang, Siqi Zheng, Yafeng Chen, Luyao Cheng, and Qian Chen.
\newblock {CAM++: A Fast and Efficient Network for Speaker Verification Using Context-Aware Masking}.
\newblock In {\em Proc. INTERSPEECH 2023}, pages 5301--5305, 2023.

\bibitem{transformer}
Ashish Vaswani, Noam Shazeer, Niki Parmar, Jakob Uszkoreit, Llion Jones, Aidan~N Gomez, {\L}ukasz Kaiser, and Illia Polosukhin.
\newblock Attention is all you need.
\newblock {\em Advances in neural information processing systems}, 30, 2017.

\bibitem{wang2021unispeech}
Chengyi Wang, Yu~Wu, Yao Qian, Kenichi Kumatani, Shujie Liu, Furu Wei, Michael Zeng, and Xuedong Huang.
\newblock Unispeech: Unified speech representation learning with labeled and unlabeled data.
\newblock In {\em International Conference on Machine Learning}, pages 10937--10947. PMLR, 2021.

\bibitem{9814838}
Sanyuan Chen, Chengyi Wang, Zhengyang Chen, Yu~Wu, Shujie Liu, Zhuo Chen, Jinyu Li, Naoyuki Kanda, Takuya Yoshioka, Xiong Xiao, Jian Wu, Long Zhou, Shuo Ren, Yanmin Qian, Yao Qian, Jian Wu, Michael Zeng, Xiangzhan Yu, and Furu Wei.
\newblock Wavlm: Large-scale self-supervised pre-training for full stack speech processing.
\newblock {\em IEEE Journal of Selected Topics in Signal Processing}, 16(6):1505--1518, 2022.

\bibitem{9747814}
Zhengyang Chen, Sanyuan Chen, Yu~Wu, Yao Qian, Chengyi Wang, Shujie Liu, Yanmin Qian, and Michael Zeng.
\newblock Large-scale self-supervised speech representation learning for automatic speaker verification.
\newblock In {\em ICASSP 2022 - 2022 IEEE International Conference on Acoustics, Speech and Signal Processing (ICASSP)}, pages 6147--6151, 2022.

\bibitem{MFA-Conformer}
Yang Zhang, Zhiqiang Lv, Haibin Wu, Shanshan Zhang, Pengfei Hu, Zhiyong Wu, Hung yi~Lee, and Helen Meng.
\newblock {MFA-Conformer: Multi-scale Feature Aggregation Conformer for Automatic Speaker Verification}.
\newblock In {\em Proc. Interspeech 2022}, pages 306--310, 2022.

\bibitem{gulati20_interspeech}
Anmol Gulati, James Qin, Chung-Cheng Chiu, Niki Parmar, Yu~Zhang, Jiahui Yu, Wei Han, Shibo Wang, Zhengdong Zhang, Yonghui Wu, and Ruoming Pang.
\newblock {Conformer: Convolution-augmented Transformer for Speech Recognition}.
\newblock In {\em Proc. Interspeech 2020}, pages 5036--5040, 2020.

\bibitem{10096659}
Danwei Cai, Weiqing Wang, Ming Li, Rui Xia, and Chuanzeng Huang.
\newblock Pretraining conformer with asr for speaker verification.
\newblock In {\em ICASSP 2023 - 2023 IEEE International Conference on Acoustics, Speech and Signal Processing (ICASSP)}, pages 1--5, 2023.

\bibitem{chung18b_interspeech}
Joon~Son Chung, Arsha Nagrani, and Andrew Zisserman.
\newblock {VoxCeleb2: Deep Speaker Recognition}.
\newblock In {\em Proc. Interspeech 2018}, pages 1086--1090, 2018.

\bibitem{SwinTransformer}
Ze~Liu, Yutong Lin, Yue Cao, Han Hu, Yixuan Wei, Zheng Zhang, Stephen Lin, and Baining Guo.
\newblock Swin transformer: Hierarchical vision transformer using shifted windows.
\newblock In {\em Proceedings of the IEEE/CVF international conference on computer vision}, pages 10012--10022, 2021.

\bibitem{10096333}
Mufan Sang, Yong Zhao, Gang Liu, John~H.L. Hansen, and Jian Wu.
\newblock Improving transformer-based networks with locality for automatic speaker verification.
\newblock In {\em ICASSP 2023 - 2023 IEEE International Conference on Acoustics, Speech and Signal Processing (ICASSP)}, pages 1--5, 2023.

\bibitem{NAT}
Ali Hassani, Steven Walton, Jiachen Li, Shen Li, and Humphrey Shi.
\newblock Neighborhood attention transformer.
\newblock In {\em Proceedings of the IEEE/CVF Conference on Computer Vision and Pattern Recognition}, pages 6185--6194, 2023.

\bibitem{dosovitskiy2020image}
Alexey Dosovitskiy, Lucas Beyer, Alexander Kolesnikov, Dirk Weissenborn, Xiaohua Zhai, Thomas Unterthiner, Mostafa Dehghani, Matthias Minderer, Georg Heigold, Sylvain Gelly, et~al.
\newblock An image is worth 16x16 words: Transformers for image recognition at scale.
\newblock In {\em ICLR}, 2020.

\bibitem{hassani2021escaping}
Ali Hassani, Steven Walton, Nikhil Shah, Abulikemu Abuduweili, Jiachen Li, and Humphrey Shi.
\newblock Escaping the big data paradigm with compact transformers.
\newblock {\em arXiv preprint arXiv:2104.05704}, 2021.

\bibitem{touvron2021training}
Hugo Touvron, Matthieu Cord, Matthijs Douze, Francisco Massa, Alexandre Sablayrolles, and Herv{\'e} J{\'e}gou.
\newblock Training data-efficient image transformers \& distillation through attention.
\newblock In {\em ICML}, 2021.

\bibitem{ramachandran2019stand}
Prajit Ramachandran, Niki Parmar, Ashish Vaswani, Irwan Bello, Anselm Levskaya, and Jon Shlens.
\newblock Stand-alone self-attention in vision models.
\newblock {\em Advances in neural information processing systems}, 32, 2019.

\bibitem{hassani2024faster}
Ali Hassani, Wen-Mei Hwu, and Humphrey Shi.
\newblock Faster neighborhood attention: Reducing the o(n\^2) cost of self attention at the threadblock level, 2024.

\bibitem{yamamoto2019speaker}
Hitoshi Yamamoto, Kong~Aik Lee, Koji Okabe, and Takafumi Koshinaka.
\newblock Speaker augmentation and bandwidth extension for deep speaker embedding.
\newblock In {\em Interspeech}, pages 406--410, 2019.

\bibitem{ko2015audio}
Tom Ko, Vijayaditya Peddinti, Daniel Povey, and Sanjeev Khudanpur.
\newblock Audio augmentation for speech recognition.
\newblock In {\em Interspeech}, volume 2015, page 3586, 2015.

\bibitem{wang2020dkudukeece}
Weiqing Wang, Danwei Cai, Xiaoyi Qin, and Ming Li.
\newblock The dku-dukeece systems for voxceleb speaker recognition challenge 2020, 2020.

\bibitem{snyder201musan}
David Snyder, Guoguo Chen, and Daniel Povey.
\newblock Musan: A music, speech, and noise corpus.
\newblock {\em arXiv preprint arXiv:1510.08484}, 2015.

\bibitem{ko2017study}
Tom Ko, Vijayaditya Peddinti, Daniel Povey, Michael~L Seltzer, and Sanjeev Khudanpur.
\newblock A study on data augmentation of reverberant speech for robust speech recognition.
\newblock In {\em 2017 IEEE international conference on acoustics, speech and signal processing (ICASSP)}, pages 5220--5224. IEEE, 2017.

\bibitem{nagrani17_interspeech}
Arsha Nagrani, Joon~Son Chung, and Andrew Zisserman.
\newblock {VoxCeleb: A Large-Scale Speaker Identification Dataset}.
\newblock In {\em Proc. Interspeech 2017}, pages 2616--2620, 2017.

\bibitem{ravanelli2021speechbrain}
Mirco Ravanelli, Titouan Parcollet, Peter Plantinga, Aku Rouhe, Samuele Cornell, Loren Lugosch, Cem Subakan, Nauman Dawalatabad, Abdelwahab Heba, Jianyuan Zhong, et~al.
\newblock Speechbrain: A general-purpose speech toolkit.
\newblock {\em arXiv preprint arXiv:2106.04624}, 2021.

\bibitem{deng2019arcface}
Jiankang Deng, Jia Guo, Niannan Xue, and Stefanos Zafeiriou.
\newblock Arcface: Additive angular margin loss for deep face recognition.
\newblock In {\em Proceedings of the IEEE/CVF conference on computer vision and pattern recognition}, pages 4690--4699, 2019.

\bibitem{deng2020sub}
Jiankang Deng, Jia Guo, Tongliang Liu, Mingming Gong, and Stefanos Zafeiriou.
\newblock Sub-center arcface: Boosting face recognition by large-scale noisy web faces.
\newblock In {\em Computer Vision--ECCV 2020: 16th European Conference, Glasgow, UK, August 23--28, 2020, Proceedings, Part XI 16}, pages 741--757. Springer, 2020.

\bibitem{sutskever2013importance}
Ilya Sutskever, James Martens, George Dahl, and Geoffrey Hinton.
\newblock On the importance of initialization and momentum in deep learning.
\newblock In {\em International conference on machine learning}, pages 1139--1147. PMLR, 2013.

\bibitem{loshchilov2016sgdr}
Ilya Loshchilov and Frank Hutter.
\newblock Sgdr: Stochastic gradient descent with warm restarts.
\newblock {\em arXiv preprint arXiv:1608.03983}, 2016.

\bibitem{cai2022kriston}
Qutang Cai, Guoqiang Hong, Zhijian Ye, Ximin Li, and Haizhou Li.
\newblock The kriston ai system for the voxceleb speaker recognition challenge 2022, 2022.

\bibitem{karam2011towards}
Zahi~N Karam, William~M Campbell, and Najim Dehak.
\newblock Towards reduced false-alarms using cohorts.
\newblock In {\em 2011 IEEE international conference on acoustics, speech and signal processing (ICASSP)}, pages 4512--4515. IEEE, 2011.

\bibitem{matejka2017analysis}
Pavel Matejka, Ondrej Novotn{\`y}, Oldrich Plchot, Lukas Burget, Mireia~Diez S{\'a}nchez, and Jan Cernock{\`y}.
\newblock Analysis of score normalization in multilingual speaker recognition.
\newblock In {\em Interspeech}, pages 1567--1571, 2017.

\end{thebibliography}

\end{document}